\documentclass[a4paper,fleqn,usenatbib,letters]{mnras}

\usepackage[T1]{fontenc}
\usepackage{ae,aecompl}

\usepackage{graphicx}	
\usepackage{amsmath}	
\usepackage{amssymb}	
\usepackage{color}	

\title[Obscured AGNs triggered after galaxy compaction]{Obscured Active Galactic Nuclei triggered in compact star-forming galaxies}

\author[Yu-Yen Chang et al.]{
Yu-Yen~Chang,$^{1}$\thanks{E-mail: yu-yen.chang@cea.fr}
Emeric~Le~Floc'h,$^{1}$
St\'ephanie~Juneau,$^{1}$
Elisabete~da~Cunha,$^{2}$
\newauthor
Mara~Salvato,$^{3}$
Francesca Civano,$^{4,5}$
Stefano~Marchesi,$^{4,5,6}$
J. M. Gabor, $^{1}$
\newauthor
Olivier~Ilbert,$^{7}$
Clotilde~Laigle,$^{8}$
H.~J.~McCracken,$^{8}$ 
Bau-Ching Hsieh, $^{9}$
\newauthor
Peter Capak$^{10,11}$
\\
$^{1}$CEA Saclay, DSM/Irfu/Service d'Astrophysique, Orme des Merisiers, F-91191 Gif-sur-Yvette Cedex, France\\
$^{2}$The Australian National University, Mt Stromlo Observatory, Cotter Rd, Weston Creek, ACT 2611, Australia\\
$^{3}$Max Planck Institut f\"ur Plasma Physik and Excellence Cluster, 85748 Garching, Germany\\
$^{4}$Harvard-Smithsonian Center for Astrophysics, 60 Garden Street, Cambridge, MA 02138, USA\\
$^{5}$Yale Center for Astronomy and Astrophysics, 260 Whitney Avenue, New Haven, CT 06520, USA\\
$^{6}$Dipartimento di Fisica e Astronomia, Universit`a di Bologna, viale Berti Pichat 6/2, 40127 Bologna, Italy\\
$^{7}$Aix-Marseille Universit\'e, CNRS, LAM (Laboratoire d'Astrophysique de Marseille) UMR 7326, 13388, Marseille, France\\
$^{8}$Sorbonne Universit\'e, UPMC Univ Paris 06, et CNRS, UNR 7095, IAP, 98b bd Arago, F-75014, Paris, France\\
$^{9}$ASIAA Sinica, AS/NTU. No. 1, Sec. 4, Roosevelt Rd., Taipei 10617, Taiwan, R.O.C.\\
$^{10}$Caltech, 1200 E. California Blvd., Pasadena, CA 91125, USA\\
$^{11}$IPAC, 1200 E. California Blvd., Pasadena, CA 91125, USA
}

\date{Accepted 2016 December 6. Received 2016 December 6; in original form 2016 July 12}

\pubyear{2016}

\begin{document}
\label{firstpage}
\pagerange{\pageref{firstpage}--\pageref{lastpage}}
\maketitle

\begin{abstract}
We present a structural study of 182 obscured Active Galactic Nuclei (AGNs)  at z\,$\leq\,$1.5, selected in the COSMOS field from their extreme infrared to X-ray luminosity ratio and their negligible emission at optical wavelengths. We fit optical to far-infrared spectral energy distributions and analyze deep HST imaging to derive the physical and morphological properties of their host galaxies.  We find that such galaxies are more compact than normal star-forming sources at similar redshift and stellar mass, and we show that it is not an observational bias related to the emission of the AGN. Based on the distribution of their $UVJ$ colors, we also argue that this increased compactness is not due to the additional contribution of a passive bulge. We thus postulate that a vast majority of obscured AGNs reside in galaxies undergoing  dynamical compaction, similar to processes recently invoked to explain the formation of compact star-forming sources at high redshift.
\end{abstract}




\section{Introduction}
\label{sec1}

The census of the population of X-ray and optically-selected Active Galactic Nuclei (AGNs) has undergone dramatic improvements over the past few years, and a general consensus on the nature of their host galaxies has now emerged : the vast majority of AGNs reside in normal isolated galaxies, in which the activity of star formation co-evolves with the super-massive black hole accretion \citep{2011ApJ...726...57C,2012ApJ...744..148K,2012MNRAS.419...95M,2013ApJ...764..176J}. However, an ongoing debate still persists on the exact structural properties of the AGN host galaxies. When compared to normal star-forming disks at $z\sim 1$, X-ray AGN hosts have been found either bulgier \citep{2005ApJ...627L..97G,2007ApJ...660L..19P}, similar \citep{2013A&A...549A..46B,2014MNRAS.439.3342V}, or intermediate between disks and spheroids \citep{2009ApJ...691..705G}.  Such inconsistent results might be partly explained by the difficulty to separate the contribution of the central AGN from the emission of the underlying host, even when high resolution images such as those obtained with the Hubble Space Telescope (HST) are used.  

Obscured AGNs weakly contribute to the total galaxy emission at optical wavelengths, and the characterization of their host is therefore less subject to the subtraction of the AGN signature. Interestingly, analysis of X-ray selected Compton-thick AGNs and their host galaxies have revealed a fraction of disturbed morphologies and mergers increasing with obscuration, substantially larger than found among moderately obscured and unobscured AGNs \citep{2015ApJ...814..104K,2015A&A...578A.120L}. Yet, most of the obscured AGNs studied so far have among the highest bolometric luminosities measured ($L_{bol}>10^{43} L_\odot$) and may not be representative of the bulk of the obscured AGN population. 

In this letter we identify AGNs from their characteristic mid-IR power-law emission using the Cosmic Evolution Survey  \citep[COSMOS,][]{2007ApJS..172....1S}  and focus on a sub-sample of obscured AGNs carefully selected based on their infrared to X-ray luminosity ratio. 
Supplementing the more traditional X-ray Compton-thick selection, our technique provides us with a large and complementary sample of obscured AGNs at moderate luminosities, allowing a more complete view on the properties of the galaxies hosting this population. We use AB magnitudes and adopt the cosmological parameters ($\Omega_M$,$\Omega_\Lambda$,$h$)=(0.35,0.70,0.70) and the \citet{2003PASP..115..763C} stellar initial mass function.


\section{The data}
\label{sec2}

\subsection{Parent galaxy sample and IR AGN candidates}

Infrared-luminous AGN candidates (hereafter IR-AGNs) were identified among the 24$\mu$m sources of the COSMOS field \citep{2009ApJ...703..222L} using a criterion similar to that initially proposed by \citet{2004ApJS..154..166L}.
In practice, we first considered all sources with IRAC F$_\nu$  flux densities monotonically rising from 3.6 to 8$\mu$m, as well as those with IRAC 5.8$\mu$m and 8$\mu$m fluxes  exceeding by $>$3$\sigma$  the expected emission of the stellar component assuming typical galaxy models and spectral energy distribution (SED) fitting techniques. 
In the parameter space defined by the $F_{5.8\mu\text{m}}/F_{3.6\mu\text{m}}$ and $F_{8\mu\text{m}}/F_{4.5\mu\text{m}}$ colors, these objects populate a specific area that shows also minimized contamination by star-forming galaxies at low to intermediate redshifts (Chang et al., in prep.). Assuming the latest photometric and spectroscopic redshifts in COSMOS  all 24$\mu$m sources at z\,$\leq\,$1.5 located in this two-color region were hence selected  as potential AGN-dominated galaxies. This led to a total sample of 655 IR-AGN candidates, 353 of which are also detected  in the deep X-ray Chandra COSMOS Legacy with an estimated $L_{2-10\text{keV}}$ luminosity greater than $10^{42}$\,erg\,s$^{-1}$ \citep{2016ApJ...817...34M,2016ApJ...819...62C}.
Among the 655 selected AGNs, 426 have reliable spectroscopy. For the other 229 sources, 106 are X-ray detected and we take the photometric redshifts from \citet{2016ApJ...817...34M} as described in \citet{2011ApJ...742...61S}.
For the remaining 123 AGNs , we assumed that the optical/NIR emission is mostly powered by the host galaxies and thus used the photometric redshifts derived by \citet{2016ApJS..224...24L}. 

\begin{figure}
\centering
\includegraphics[width=1.0\columnwidth,height=0.293\textwidth]{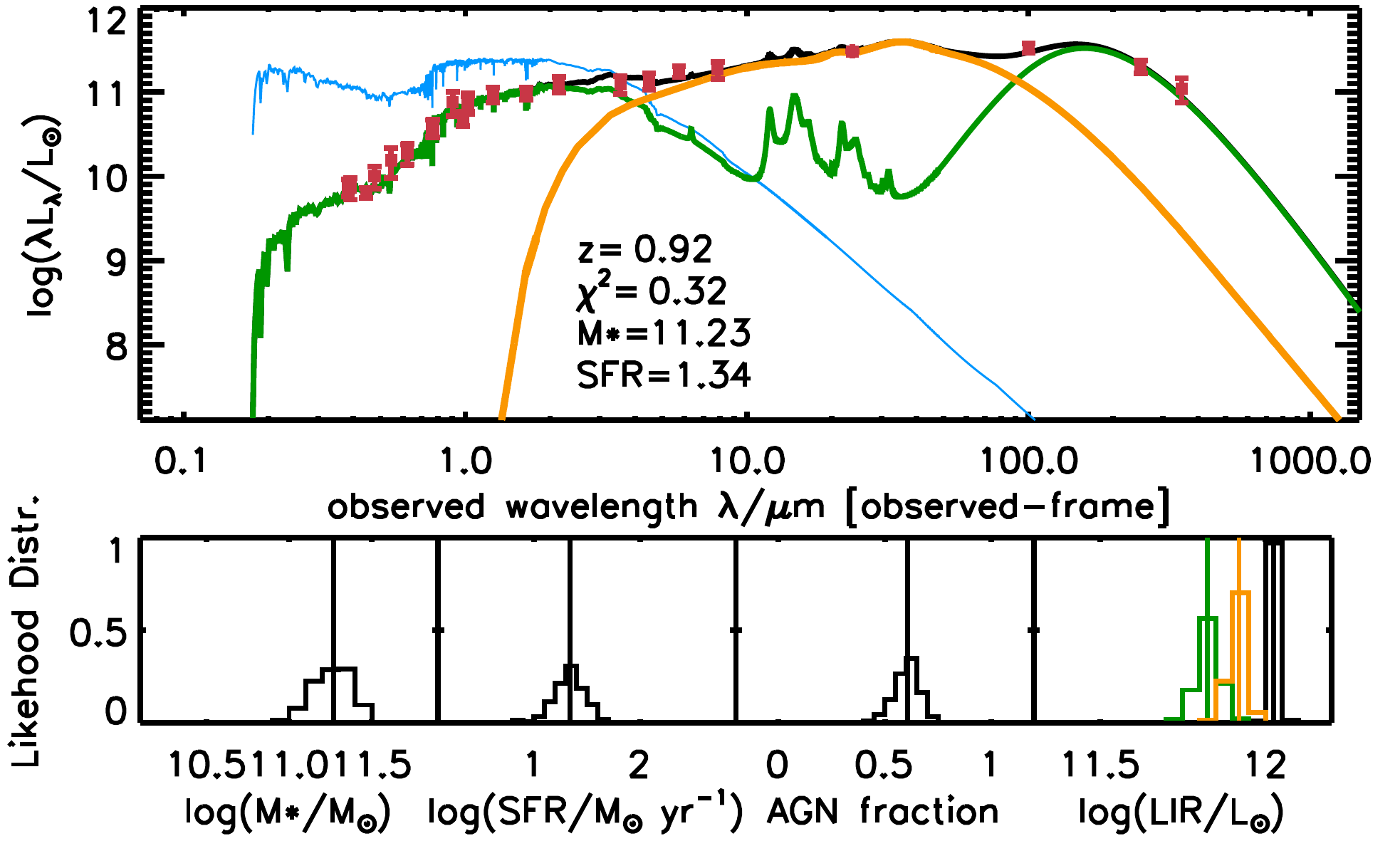} 
\caption[]{The multi-band photometry of a type-2 IR-AGN (top panel, red data points) along with its best SED fit (black solid line). The emission from the stellar component and the AGN are represented by the green and orange curves respectively, while the blue line shows the unattenuated stellar spectrum. Likelihood distributions are illustrated for the stellar mass, the SFR, AGN fraction (contribution to the infrared), and IR luminosity.}
\label{cosl_sed}
\end{figure}

All COSMOS 24$\mu$m sources have optical to far-IR photometry available from the band-merged COSMOS2015 catalog of \citet{2016ApJS..224...24L}. We fit SEDs to the IR-AGNs with a new version of MAGPHYS \citep[in prep.]{2008MNRAS.388.1595D,2015ApJ...806..110D}, updated to account for a possible AGN contribution and relying on both obscured and unobscured AGN templates.
Fig.\,~\ref{cosl_sed} shows the multi-band photometry of one IR-AGN candidate and its host galaxy, along with the best SED fit decomposed into the relative AGN and stellar contributions. We find that galaxies hosting IR-AGNs at $z$\,$\leq$\,1.5 in COSMOS have typical stellar masses M$_*$\,$\gtrsim$\,10$^{10}$\,M$_\odot$ with a peak distribution at M$_*$\,$\sim$\,10$^{10.5}$\,M$_\odot$. 
They have specific Star Formation Rates (sSFR) typical of star-forming sources, 
45\% of them being individually detected with {\it Herschel} in the far-IR, i.e., in a wavelength range where the dust emission is mostly powered by intense star formation.

\begin{figure*}
\centering
\includegraphics[width=0.45\textwidth,height=0.365\textwidth]{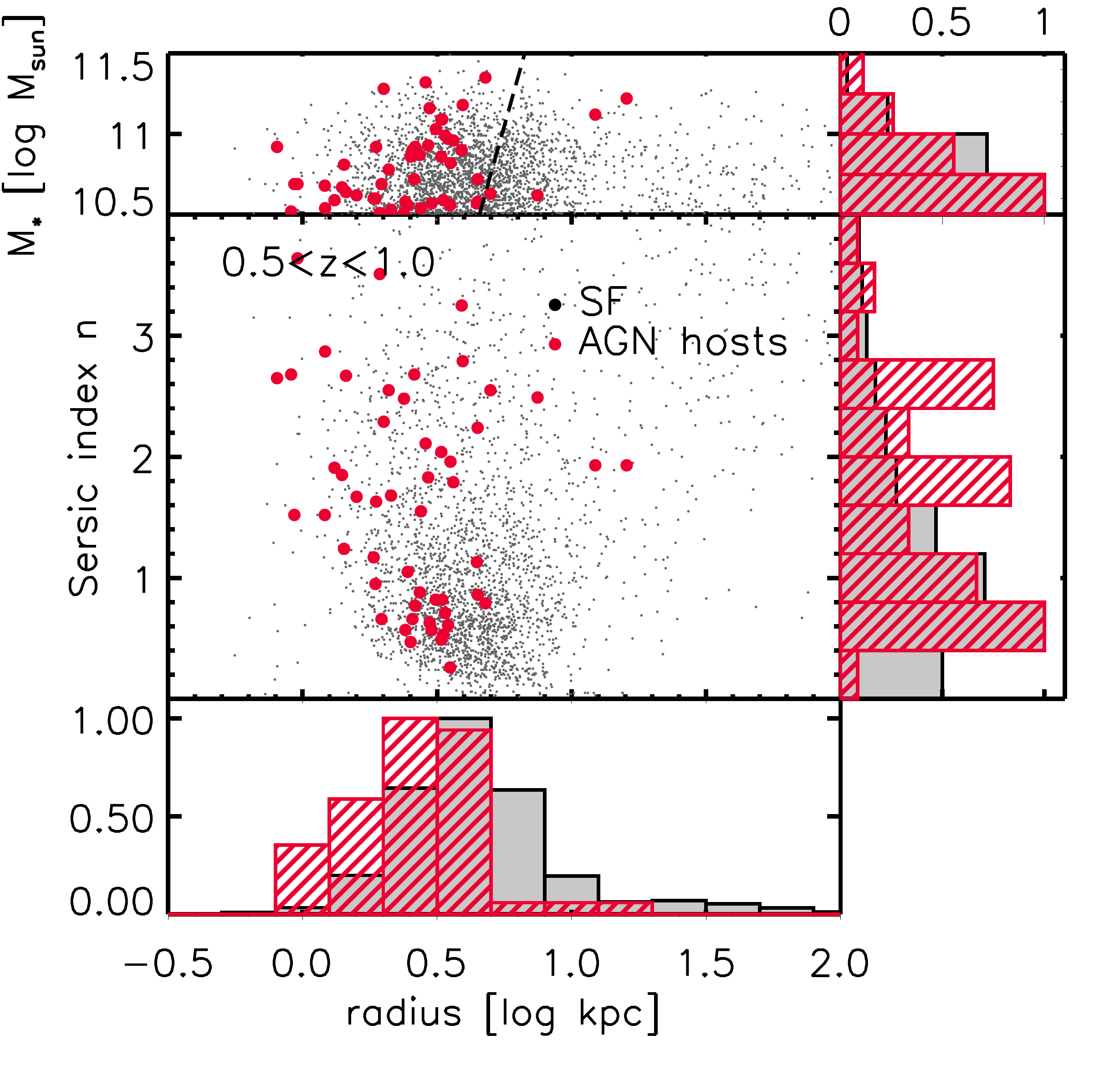}
\includegraphics[width=0.45\textwidth,height=0.365\textwidth]{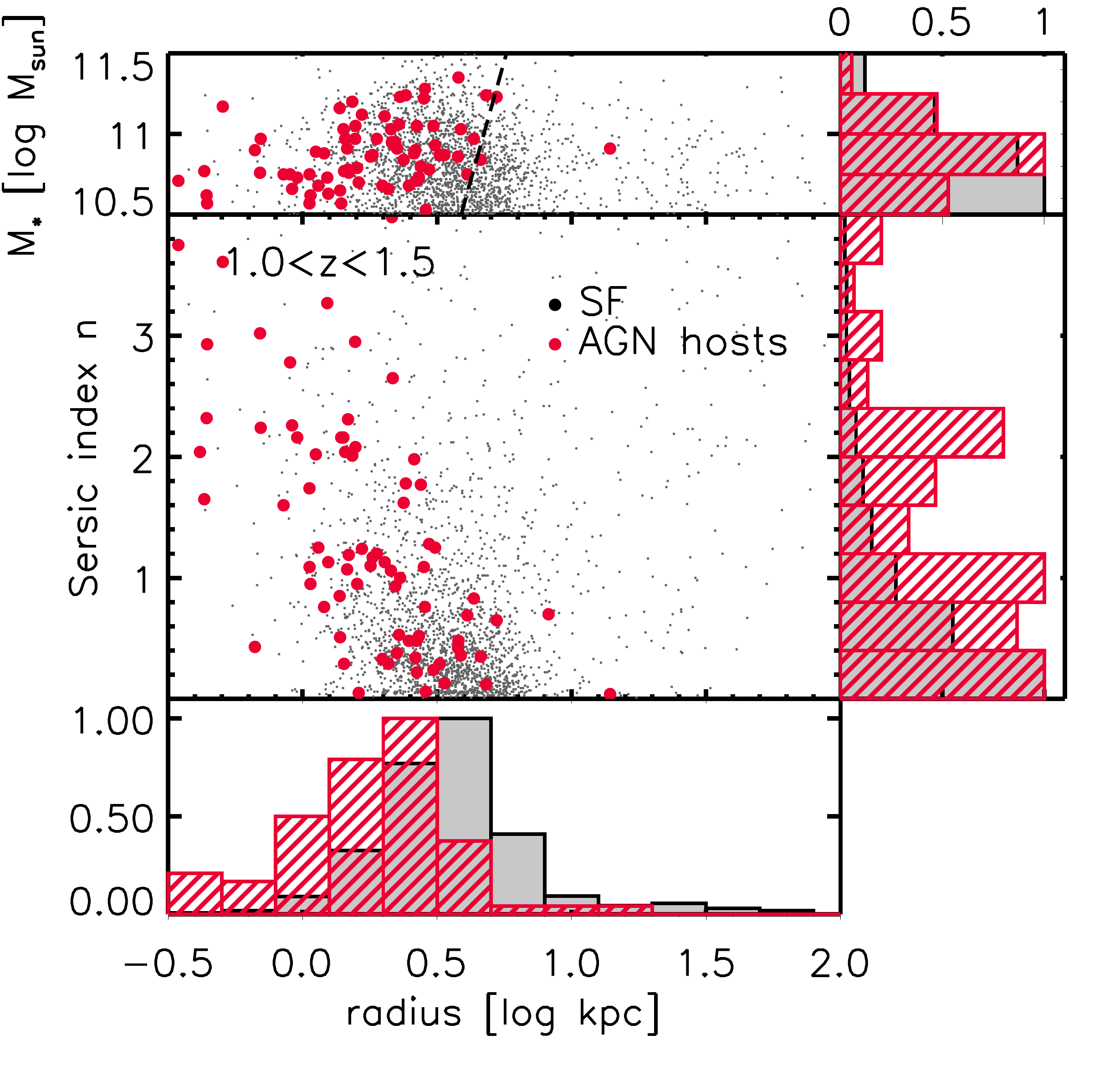} 
\caption[]{Stellar mass and S\'ersic index versus half-light radius of obscured AGN hosts (red dots), compared to star-forming galaxies (black) at $0.5<z<1$ (left) and $1<z<1.5$ (right). The corresponding distributions are represented with normalized histograms along  the $x$- and $y$-axis. The hosts of obscured AGNs have smaller radii and higher S\'ersic indices than star-forming sources. The dashed-line indicates the mass--size relation of star-forming galaxies reported by \citet{2014ApJ...788...28V}.
}
\label{cosl_rn}
\end{figure*}

\subsection{Obscured AGNs}

Assuming the total infrared luminosity of the AGN component derived from the SED fits (referred as $L_{IR,  AGN}$ hereafter)
we define ``obscured AGNs" as IR-AGNs with $L_{IR,  AGN} / L_X > 20$, where $L_X$ is the X-ray 2-10 keV luminosity or upper limit obtained from the Chandra Legacy observations of COSMOS. 
For the latter we assumed $\Gamma = 1.4$ following the prescriptions of \citet{2016ApJ...817...34M}. We used the 90\% completeness flux limit $f_{2-10\text{keV}}\,=\,5.0\times10^{-15}$\,erg\,s$^{-1}$\,cm$^{-2}$ in case of no X-ray detection, resulting in a selection of 265 objects.
Given the relation $L_{IR}$\,$\sim$\,2\,$\times$\,$L_{6\mu\text{m}}$ implied by our type-2 AGN SED models, we note that our criterion identifies AGNs with dust reprocessed emission at least 0.3\,dex larger than expected from the intrinsic $L_{6\mu\text{m}} - L_X$ relationship observed among unobscured AGNs \cite[e.g.,][]{2015ApJ...807..129S}. Therefore, our selection is not restrictive to extreme levels of X-ray absorption (e.g., a Compton-thick AGN with $N_H$\,=\,$10^{24}$ cm$^{-2}$ would correspond to $L_{IR} / L_X \sim 80 $, \citet{2009ApJ...693..447F}) but it provides  a statistically large sample of  AGNs  with high levels of obscuration, among which 
$\sim$\,50\% escape the standard X-ray selection criterion  L$_X$\,$>$\,10$^{42}$ erg s$^{-1}$.

Because of small-scale dust-free gas absorbers within the broad line region, AGNs with strong obscuration in the X-rays can still contribute some light at UV/optical wavelengths \citep[e.g.,][]{2013MNRAS.431..978L,2014MNRAS.437.3550M}.
To avoid skewing our morphology measurements, 34 sources with detectable AGN contribution in the optical were excluded from our sample: based on our SED fitting, we only kept sources where the AGN is best described by a deeply obscured torus emission contributing less than 0.001\% to the I-band, ensuring that the optical light is fully dominated by the stellar component of the host galaxy such as in Fig.1 
{\footnote{We made the stacking of the Probability Distribution Functions of the AGN component resulting from the decomposition of each source in our sample and confirm a very high probability (>96.95\%) to host a deeply buried AGN with no detectable emission (<0.001\%) in the $I$-band.}
Independently of our SED fits, we also removed 17 AGNs previously  identified as type-1's by \citet{2011ApJ...742...61S} 
and \citet{2009ApJS..184..218L}.
Restricting our analysis to the 1.64\,deg$^2$ HST/ACS imaging of COSMOS, we were left with a final sample of 182 obscured AGNs at z\,$\leq\,$1.5.


\section{Morphology}
\label{sec3}

We analyzed the morphology of the obscured AGN host galaxies using the $I$-band ACS imaging of COSMOS, 
with both a visual classification approach and a parametric two-dimensional surface brightness fitting method. While the visually-classified properties will be reported in a future publication, we discuss hereafter the results  obtained by fitting the $I$-band light distribution 
with the GALFIT algorithm \citep{2010AJ....139.2097P}. 
Having eliminated all the sources with detectable AGN contribution and avoiding degeneracy from multiple components, it is fair to assume that the HST image can be fitted by using a S\'ersic profile only as it is confirmed by radial light profile later.
Fig.\,~\ref{cosl_rn} shows the distributions of their half-light radius and Sersic index measured for M$_*$\,$>$\,10$^{10.5}$\,M$_\odot$ at $0.5<z<1$ and $1<z<1.5$, compared to the distributions obtained for a control sample of star-forming galaxies selected from the COSMOS2015 catalog \citep{2016ApJS..224...24L} at similar redshifts and comparable stellar masses. 

It is clear that the host galaxies of obscured AGNs have on average smaller radii and larger S\'ersic indices, a result confirmed at $>$\,99\% confidence by a Kolmogorov-Smirnov (K-S) test. 
These galaxies could thus share some properties with the hosts of X-ray bright and radio-selected AGNs, which also seem to be more bulgy than star-forming galaxies even after subtracting the emission from the AGN \citep{2009ApJ...691..705G,2016MNRAS.458.2391B}. Similarly, they appear to be shifted toward smaller sizes with respect to the mass--size relation of star-forming galaxies \citep{2014ApJ...788...28V}.

To further quantify this trend, 
we investigated the averaged galaxy radial light profile using a stacking analysis. For each object, an ACS image centered on the peak emission of the best-fit two-dimensional  model given by GALFIT was extracted and re-sampled to a common physical scale of 0.18\,kpc/px using bilinear interpolation. The sample of star-forming galaxies was matched in stellar mass to the AGNs, and an averaged stack was created separately for the two populations in the two redshift bins mentioned above.  Fig.\,~\ref{cosl_profile} shows the corresponding stacked images along with the average radial profile of the associated stacked detection, as well as the structural parameters of the best fits obtained with GALFIT. Here again, the $I$-band light distribution characterizing the hosts of obscured AGNs appears more compact than the profile of star forming galaxies. If we further sub-divide and compare the two populations by bin of galaxy stellar mass, the same trend is still observed for each redshift range, albeit with a lower significance at M$_*$\,$\gtrsim$\,10$^{11}$\,M$_\odot$ probably altered by the lower signal to noise at the most massive end due to small sample size.

\begin{figure*}
\centering
\includegraphics[width=0.80\textwidth]{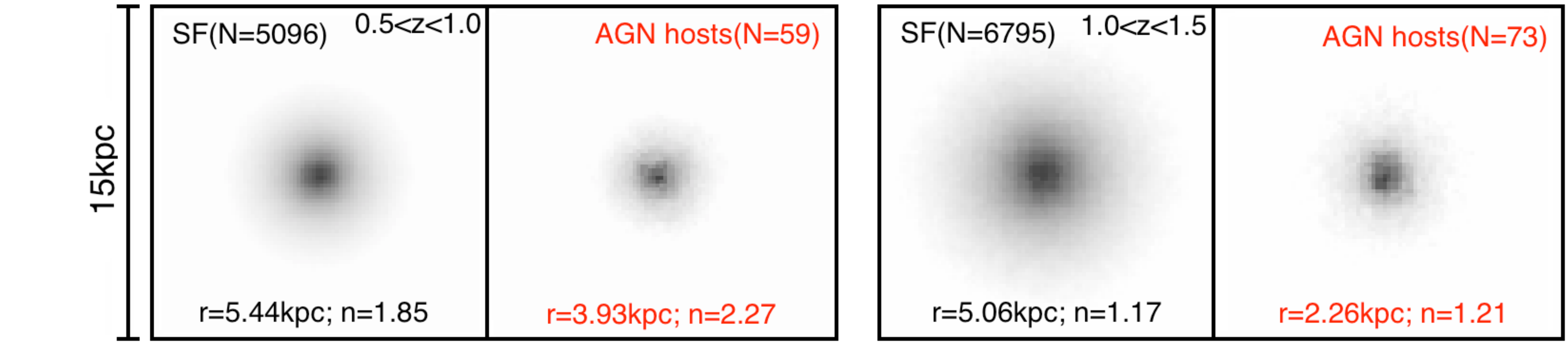} 
\includegraphics[width=0.80\textwidth]{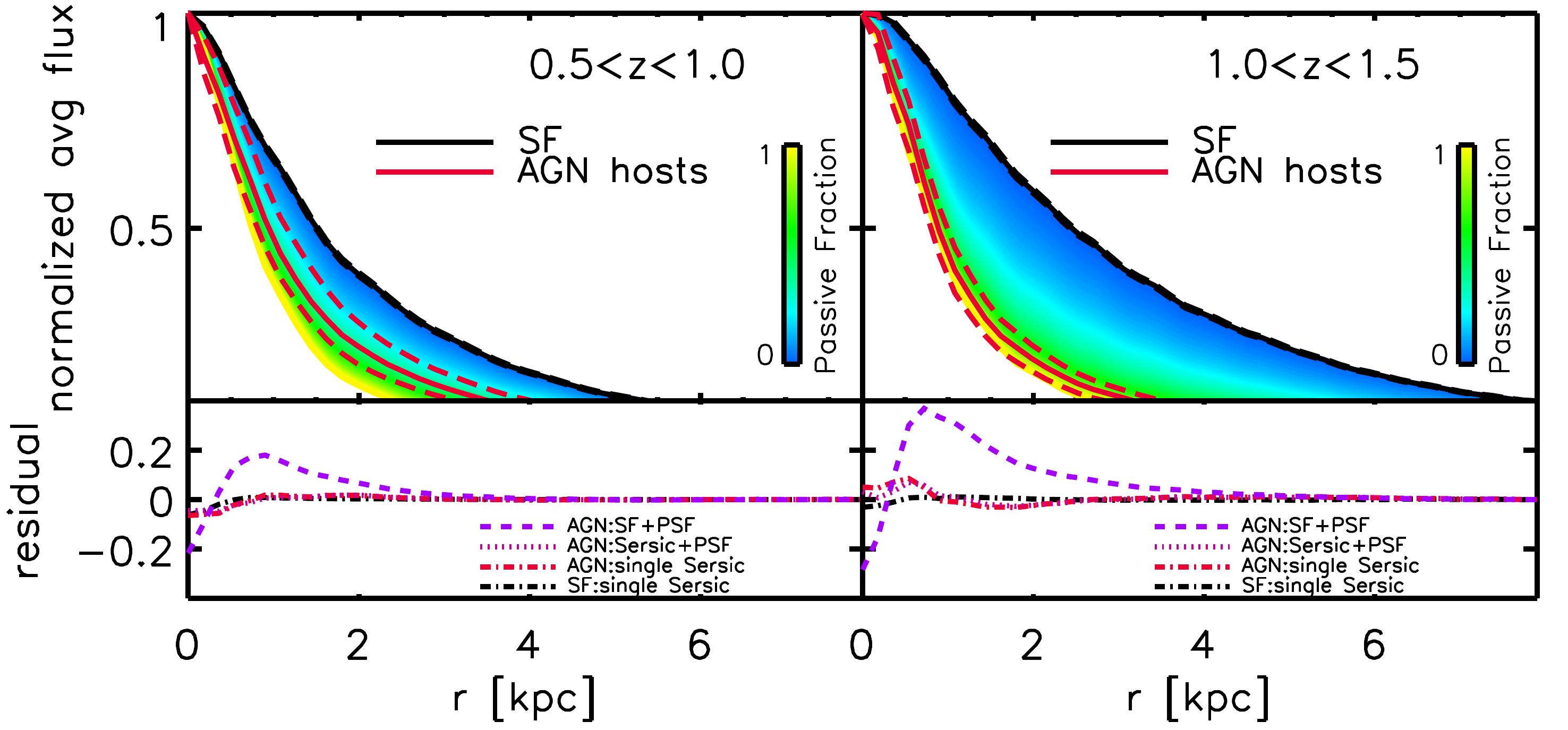} 
\caption[]{{\it  Top:} Normalized stacked images of obscured AGN hosts and star-forming sources, displayed with a common grey-scale and selected as in Fig.\,\ref{cosl_rn}, along with the half-light radius and the S\'ersic index measured with GALFIT. {\it Bottom:} Average radial light profile obtained from the  stacked images (red: obscured AGN hosts; black: star-forming sources). Solid  and dashed lines represent the median profile and the 1$\sigma$ dispersion estimated with bootstrapping. 
 The colored regions depict the radial profiles obtained after combining the stack of star-forming galaxies with the stack of $UVJ$-selected red passive sources, with the contribution of the latter color-coded as shown in the inset.Dash-dotted and dotted curves illustrate the residuals after fitting the stacks with a  single S\'ersic disk and a S\'ersic+PSF two-component model, respectively. The blue dashed line shows the residuals after fitting the AGN stack with a PSF and a S\'ersic disk fixed to that obtained for the population of star-forming sources.
}
\label{cosl_profile}
\end{figure*}

\begin{figure}
\centering
\includegraphics[width=0.40\textwidth,height=0.30\textwidth]{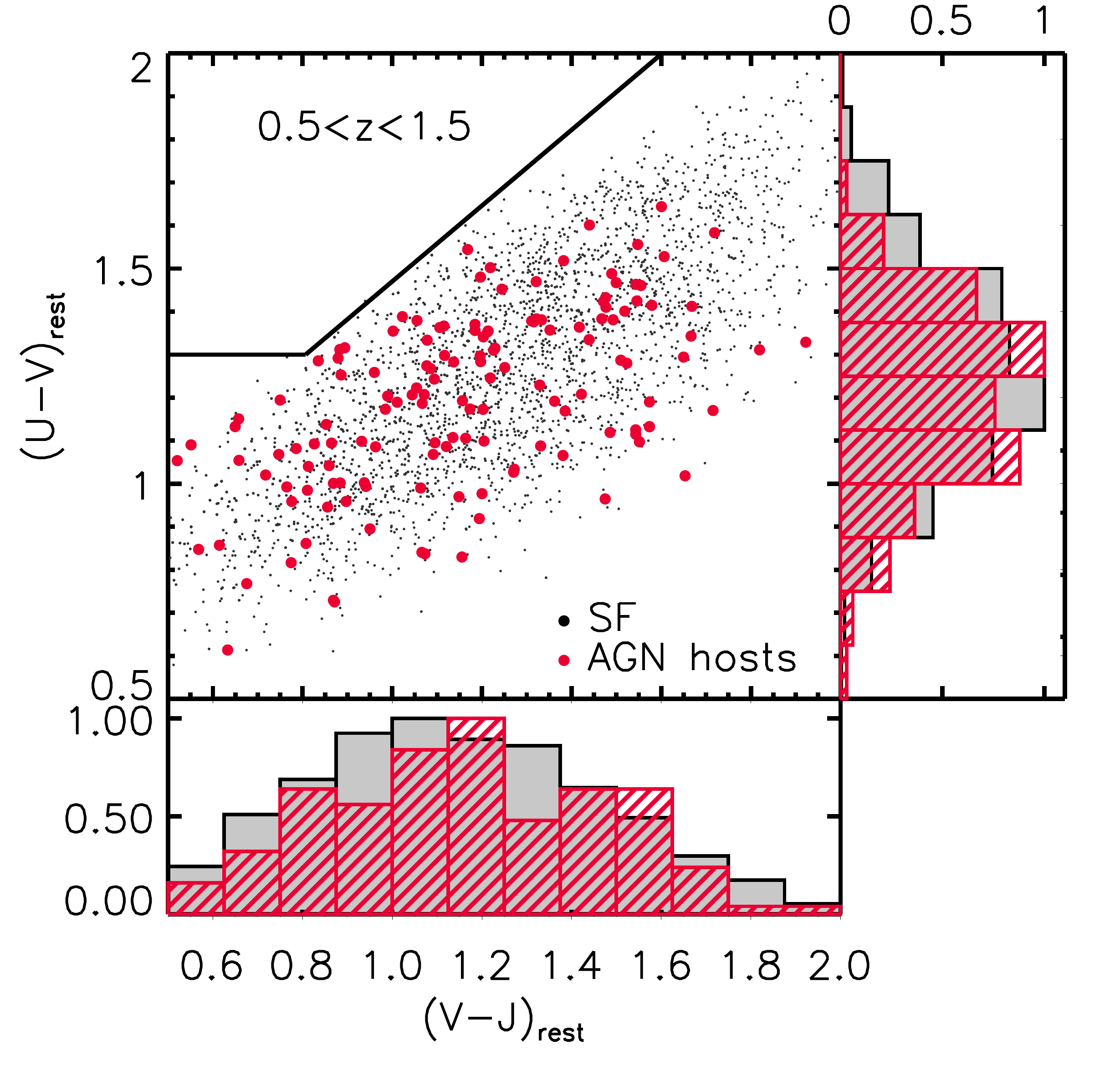} 
\caption[]{The $UVJ$ rest-frame colors of AGN hosts (red dots, stellar component only) compared to the control sample of star-forming galaxies. The corresponding  normalized histograms are shown along  the $x$- and $y$-axis. The hosts of obscured AGNs have optical/NIR colors typical of star-forming sources. 
}
\label{cosl_uvj}
\end{figure}

We stress that this steeper light profile of AGN hosts can not be due to the impact of the central AGN nor to the presence of a compact circum-nuclear starburst. Considering that in such cases the emission would be unresolved at the resolution of HST, we used GALFIT to analyse the AGN stacked images assuming the combination of a S\'ersic disk and a PSF. 
Fixing the geometry of the first component to that obtained for the control sample of star-forming sources and adopting a PSF does not lead to acceptable fits. This profile overpredicts the central flux while decreasing too steep on small scales (<1kpc).
Similarly, leaving the S\'ersic index and the radius as free parameters results in a small PSF fraction ($\lesssim$\,3.5\%) and thus a disk component again much steeper than the average profile of star-forming galaxies (see Fig.\,\ref{cosl_profile}). The more compact profile we observe among obscured AGNs is therefore intrinsically related to the morphology of their hosts, suggesting either a global compaction of the star-forming  disk  or the presence of an additional compact core in the central region. 

We tested the latter option with the hypothesis that an old stellar bulge has already formed in these objects. We considered the COSMOS red passive sources selected at similar redshift and stellar mass using the $nUVrJ$ color-color criterion from \citet{2016ApJS..224...24L} and derived their average radial light profile as described above. As illustrated in Fig.\,\ref{cosl_profile}, we find that the total light profile of obscured AGN hosts can not be reliably reproduced by combining a  red stellar bulge with the typical disk component characterizing the star-forming galaxies of the field (color-coded swath). Furthermore, the stack modeling suggests that the contribution of a possible red bulge to the total $I$-band light should be larger than $\sim$\,60\% in order to steepen the radial profile as observed, which is in strong disagreement with the color distribution of AGN host galaxies at optical wavelengths. For instance, we show in Fig.\,\ref{cosl_uvj} the standard $UVJ$ rest-frame color plot obtained for  the hosts of obscured AGNs (assuming the best-fit stellar component derived from our SED fitting), compared to the control sample of star-forming galaxies. Assuming a typical  passive galaxy SED, a $\sim$\,60\% bulge contribution  to the $I$-band light would result in  $U-V$ colors statistically redder for the AGN hosts,
while the two samples exhibit remarkably similar color distributions. We thus conclude that the apparent compactness of obscured AGN host galaxies is  not produced by the  additional  contribution of an old stellar bulge. 


\section{Discussion}
\label{sec4}

Obscured AGNs  have negligible emission at UV/optical wavelengths, bringing the benefit of a minimal impact when characterizing the morphology of their underlying hosts with facilities like HST. 
Here we have shown that these hosts exhibit a radial light profile on average more compact than that of star-forming sources at similar mass, but not as steep as that observed in red passive objects (Fig.\,\ref{cosl_profile}). At very first sight, this could be interpreted as the signature of  ``transition galaxies" migrating from the population of disk-dominated blue sources to the red sequence of dead ellipticals, with the AGN possibly 
holding a negative feedback action responsible for the quenching of the star-forming activity. 
Numerical simulations indeed suggest that wet major mergers can trigger a burst  of intense star formation accompanied by accretion of material onto a highly obscured  black hole, eventually leading to the formation of a passive galaxy once the gas has been blown out \citep[e.g.,][]{2008ApJS..175..356H}.
In this scenario though, we expect the obscured AGN to be observed only in the very first stages of the merger before the formation of the compact remnant and when the morphology is still highly perturbed. This  is not consistent with the low fraction of mergers observed in our sample (Chang et al. in prep.), which also  gives further support to the now widely-accepted scheme in which the bulk of AGNs is {\it not\,} primarily linked to galaxy mergers. 
Besides, obscured  AGN hosts have optical/NIR colors (Fig.\,\ref{cosl_uvj}) and sSFRs  typical of star-forming galaxies
(Chang et al. in prep.), which is hard to reconcile with sources in a quenching phase.

More likely,  we suggest that obscured AGNs are  primarily hosted by star-forming galaxies having undergone a process of {\it  dynamical contraction},  similar to the scenario recently proposed to explain the population of blue compact star-forming sources observed at $z$\,$\sim$\,1--2. Cosmological hydrodynamical simulations of galaxy formation indeed suggest that highly perturbed gas-rich disks fed by cold streams at high redshift can experience phases of dissipative contraction, leading to the formation of compact star-forming systems \citep{2014MNRAS.438.1870D}. In this scenario, the compaction is initially triggered by an episode of intense gas inflow involving counter-rotating streams or recycled gas as well as minor mergers. The inflow provides sufficient energy for maintaining a high level of turbulence and it is often associated with violent disk instabilities that drive gas to the central region of galaxies. It thus leads to a massive core with high gas fraction and enhanced star formation \citep{2015MNRAS.450.2327Z}, and the inflow rate can also sustain gas accretion onto the central super-massive black hole leading to the trigger of an AGN with moderately sub-Eddington luminosities \citep{2011ApJ...741L..33B}. 
These disk instabilities can actually result in substantial central gas column densities with a peak distribution at log(N$_{\rm HI})$\,$\sim$\,10$^{23}$\,cm$^{-2}$ \citep{2011ApJ...741L..33B}. Gas compaction could thus lead to 
central extinctions with even larger values, consistent with the AGNs being highly obscured when they are hosted by compact star-forming sources.

Interestingly, it was found that compact star-forming sources at z\,$\sim$\,2 seem more likely to host an X-ray bright AGN when compared to more extended sources \citep{2014ApJ...791...52B}, and that both clumpy and compact disk galaxies at $z\sim2$ harbor a high AGN fraction \citep{2014ApJ...793..101T}. Here we analyzed a {\it complete\,} sample of AGNs with strong obscuration (defined by their X-ray to total infrared AGN luminosity ratio) at 0.5\,$\leq$\,z\,$\leq$\,1.5, showing that there could be  an even stronger connection between the mechanisms driving obscured black hole accretion and compact galaxies, at least at  these lower redshifts. 
Exploring further how this link is driven by the galaxy internal processes will yet require a larger sample of AGNs, which will be achieved thanks to wider AGN surveys like those to be performed by eROSITA.


\section*{Acknowledgements}

We thank the  referee as well as A. Dekel and D. Elbaz for helpful comments and discussions. We acknowledge the contribution of the entire COSMOS collaboration as well as a financial support from ANR (\#ANR-12-JS05-0008-01).




\bibliographystyle{mnras}


\bsp	
\label{lastpage}
\end{document}